\documentclass[12pt]{article}
\usepackage[english,german,french,polish]{babel}
\usepackage[T1]{fontenc}
\usepackage{amsfonts}

\begin{document}

\selectlanguage{english}

\baselineskip 0.77cm
\topmargin -0.6in
\oddsidemargin -0.1in

\let\ni=\noindent

\renewcommand{\thefootnote}{\fnsymbol{footnote}}

\newcommand{\SM}{Standard Model }

\pagestyle {plain}

\setcounter{page}{1}

\pagestyle{empty}

~~~

\begin{flushright}
IFT-- 06/11
\end{flushright}

\vspace{0.4cm}

{\large\centerline{\bf Constructing the off-diagonal part of active-neutrino mass matrix}}

{\large\centerline{\bf from annihilation and creation matrices}}

{\large\centerline{\bf in neutrino-generation space{\footnote{Work supported in part by the Polish Ministry of Higher Education and Science, grant 1 PO3B 099 29 (2005-2007). }}}} 

\vspace{0.4cm}

{\centerline {\sc Wojciech Kr\'{o}likowski}}

\vspace{0.3cm}

{\centerline {\it Institute of Theoretical Physics, Warsaw University}}
 
{\centerline {\it Ho\.{z}a 69,~~PL--00--681 Warszawa, ~Poland}}

\vspace{0.5cm}

{\centerline{\bf Abstract}}

\vspace{0.2cm}

The off-diagonal part of the active-neutrino mass matrix is constructed from two $3\times 3$ 
matrices playing the role of annihilation and creation matrices acting in the neutrino-generation 
space of $\nu_e , \nu_\mu , \nu_\tau$. The construction leads to a new relation, $M_{\mu\,\tau} = 
4\sqrt{3} M_{e\,\mu}\,$, which {\it predicts} in the case of tribimaximal neutrino mixing that $m_3 - m_1 = \eta \,(m_2 - m_1)$ with $\eta = 5.28547$. Then, the maximal possible value of ${\Delta m^2_{32}}/{\Delta m^2_{21}}$ is equal to $\eta^2 -1 = 26.9362$ and gives $m_1 = 0$. With the experimental estimate ${\Delta m^2_{21}}\sim 8.0\times 10^{-5}\;{\rm eV}^2$, this maximal value, if realized,  {\it predicts} $\Delta m^2_{32} \sim 2.2\times 10^{-3}\;{\rm eV}^2$, near to the popular experimental estimation $\Delta m^2_{32} \sim 2.4\times 10^{-3}\;{\rm eV}^2$.

\vspace{0.5cm}

\ni PACS numbers: 12.15.Ff , 14.60.Pq  .

\vspace{0.6cm}

\ni June 2006  

\vfill\eject

~~~
\pagestyle {plain}

\setcounter{page}{1}

As is well known, the so-called tribimaximal mixing matrix [1]

%rownanie 1
\begin{equation}
U = \left( \begin{array}{rrr} \frac{\sqrt2}{\sqrt3} & \frac{1}{\sqrt3} & 0\, \\ - \frac{1}{\sqrt6} & \frac{1}{\sqrt3} & \frac{1}{\sqrt2} \\ \frac{1}{\sqrt6} & -\frac{1}{\sqrt3} & \frac{1}{\sqrt2}  \end{array} \right) 
\end{equation}

\vspace{0.2cm}

\ni describes reasonably well the active-neutrino mixing

%rownanie 2
\begin{equation}
\nu_\alpha  = \sum_i U_{\alpha i}\, \nu_i \;\;(\alpha = e, \mu, \tau\,,\, i = 1,2,3)
\end{equation}

\ni in all confirmed neutrino oscillation experiments [2]. Here, $c_{12} = \sqrt{2/3}$, $s_{12} = 1/\sqrt{3}$, $c_{23} = 1/\sqrt{2} = s_{23}$ and $s_{13} = 0$. If the charged-lepton mass matrix is diagonal, the neutrino mixing matrix $U = (U_{\alpha i})$ is at the same time the diagonalizing matrix for the active-neutrino mass matrix $M = (M_{\alpha \beta})$ whose elements

%rownanie 3
\begin{equation} 
M_{\alpha\,\beta} = \sum_i U_{\alpha i}m_i U^*_{\beta i} \;\;(\alpha , \beta = e, \mu, \tau)
\end{equation}

\ni get explicitly the form

\vspace{-0.2cm}

%rownanie 4
\begin{eqnarray}
M_{e\,e} & = & \:\:\:\frac{1}{3}(2m_1 + m_2)\,, \nonumber \\
M_{\mu\,\mu} & = & \:\:\,M_{\tau\,\tau} = \:\:\:\,\frac{1}{6} (m_1 + 2m_2 + 3 m_3)\,, \nonumber \\
M_{e\,\mu} & = & -M_{e\,\tau} =  -\frac{1}{3} (m_1 - m_2)\,, \nonumber \\
M_{\mu\,\tau} & = & -\frac{1}{6} (m_1 + 2m_2 - 3 m_3)
\end{eqnarray}

\ni with $m_i$ denoting the active-neutrino masses. Thus,

%rownanie 5
\begin{equation} 
m_1 = M_{e\,e} - M_{e\,\mu} \;,\; m_2 = M_{e\,e} +2M_{e\,\mu} \;,\; m_3 = M_{\mu\,\mu} + M_{\mu\,\tau} \;.
\end{equation}

\ni In the present paper, we will use the mass matrix (4), valid in the case of tribimaximal neutrino mixing, as a reasonable approximation.

Recently, we have proposed for active neutrinos of three generations $i = 1,2,3$ the following empirical mass formula [3]:

%rownanie 6
\begin{equation}
m_{i}  = \mu \, \rho_i \left[1- \frac{1}{\xi} \left(N^2_i + \frac{\varepsilon -1}{N^2_i}\right)\right] \;\;(i = 1,2,3)\;
\end{equation}

\vspace{-0.2cm}

\ni or, rewritten explicitly,  
 
\vspace{-0.2cm}

%rownanie 7
\begin{eqnarray}
m_1 & = & \frac{\mu}{29} (1 - \frac{\varepsilon}{\xi}) \,, \nonumber \\
m_2 & = & \frac{\mu}{29}\, 4\left[1 -\frac{1}{9\xi}(80 + \varepsilon)\right] \,, \nonumber \\
m_3 & = & \frac{\mu}{29} \,24\left[ 1 -\frac{1}{25\xi}\,(624 + \varepsilon)\right] \,.
\end{eqnarray}

\vspace{0.1cm}

\ni Here, $\mu >0\,,\, \varepsilon >0$ and $\xi >0$ are three free parameters, while

%rownanie 8
\begin{equation}
N_1 = 1 \;,\; N_2 = 3 \;,\; N_3 = 5
\end{equation}

\ni and

\vspace{-0.2cm}

%rownanie 9
\begin{equation}
\rho_1 = \frac{1}{29} \;,\; \rho_2 = \frac{4}{29} \;,\; \rho_3 = \frac{24}{29}
\end{equation}

\vspace{0.1cm}

\ni ($\sum_i \rho_i = 1$). The latter numbers have been called generation-weighting factors. The empirical mass formula (6) can be supported by an intuitive model of formal intrinsic interactions which might work within leptons and quarks [4].

For normal hierarchy of neutrino masses $m^2_{1} \ll m^2_{2} \ll m^2_{3}$, when taking the lowest mass lying in the range 
 
\vspace{-0.2cm}

%rownanie 10
\begin{equation}
m_1 \sim(0 \;{\rm to}\;10^{-3})\;{\rm eV} \;,
\end{equation}

\ni we obtain from the popular experimental estimates [2]

%rownanie 11
\begin{equation}
|m^2_2 - m^2_1| \sim 8.0\times 10^{-5}\; {\rm eV}^2 \;,\; |m^2_3 - m^2_2| \sim 2.4\times 10^{-3}\; {\rm eV}^2 \;,
\end{equation}

\ni two higher masses

%rownanie 12
\begin{equation}
m_2 \sim (8.9 \;\;{\rm to}\;\; 9.0)\times 10^{-3}\; {\rm eV} \;,\; 
m_3 \sim 5.0\times 10^{-3}\; {\rm eV} \;.
\end{equation}

\vspace{0.2cm}

\ni Then, we can determine the following parameter values in Eq. (6):

%rownanie 13
\begin{equation}
\mu \sim (7.9\;{\rm to} \;7.5)\times 10^{-2}\;{\rm eV} \;,\;\frac{\varepsilon}{\xi} \sim (1\;{\rm to} \;0.61) \;,\; \frac{1}{\xi} \sim (8.1\;{\rm to} \;6.9)\times 10^{-3}\;.
\end{equation}

\ni So, the parameter $1/\xi $ in Eq. (6) is small {\it versus} 1 and $\varepsilon/\xi $.

One may try to conjecture that in Eq. (6) $1/\xi = 0$ exactly [3]. Then, one predicts $m_3 = (6/25)(27 m_2 - 8 m_1)$, implying from the estimates (11) the inverse order of $m_1$ and $m_2$: $m_1 \sim 1.5\times 10^{-2}$ eV, $m_2 \sim 1.2\times 10^{-2}$ eV, $m_3 \sim 5.1\times 10^{-2}$ eV. In this case, $\mu \sim 4.5\times 10^{-2}$ eV and $\varepsilon/\xi \sim -8.8$.

Instead, we will try in the present paper to relate the empirical mass formula (6) to the structure of 
active-neutrino mass matrix defining also the neutrino mixing. To this end, let us introduce the matrices [5]

\vspace{0.1cm}

%rownanie 14
\begin{equation} 
N  = 2 n +{\bf 1} = \left( \begin{array}{ccc} 1 & 0 & 0 \\ 0 & 3 & 0 \\ 0 & 0 & 5 \end{array}\right)\;,\; n = \left( \begin{array}{ccc} 0 & 0 & 0 \\ 0 & 1 & 0 \\ 0 & 0 & 2 \end{array}\right)\;,\; {\bf 1} = \left( \begin{array}{ccc} 1 & 0 & 0 \\ 0 & 1 & 0 \\ 0 & 0 & 1 \end{array}\right)
\end{equation}

\vspace{0.2cm}

\ni and

\vspace{-0.2cm}

%rownanie 15
\begin{equation} 
a = \left( \begin{array}{ccc} 0 & 1 & 0 \\ 0 & 0 & \sqrt2 \\ 0 & 0 & 0 \end{array}\right)  \; ,\;a^\dagger = \left( \begin{array}{ccc} 0 & 0 & 0 \\ 1 & 0 & 0 \\ 0 & \sqrt2 & 0 \end{array}\right)\;,  
\end{equation}

\ni where

\vspace{0.2cm}

%rownanie 16
\begin{equation} 
[a\, , \,n] = a \;,\; [a^\dagger\, , \,n] = -a^\dagger\;, \;n = a^\dagger a \;,\; a^3 = 0 \;,\; a^{\dagger \,3}= 0
\,,  
\end{equation}

\vspace{0.1cm}

\ni the latter, $a$ and $a^\dagger$, playing the role of annihilation and creation $3\times 3$ matrices, although

\vspace{0.1cm}

%rownanie 17
\begin{equation} 
[a\,,\, a^\dagger] = {\bf 1} - \left( \begin{array}{ccc} 0 & 0 & 0 \\ 0 & 0 & 0 \\ 0 & 0 & 3 \end{array}\right)  \neq {\bf 1}\;.
\end{equation}

\vspace{0.2cm}

\ni Note that

%rownanie 18
\begin{equation} 
a^2 = \left( \begin{array}{ccc}  0 & 0 & \sqrt2 \\ 0 & 0 & 0 \\ 0 & 0 & 0 \end{array}\right)  \; ,\;a^{\dagger\,2} = \left( \begin{array}{ccc} 0 & 0 & 0 \\ 0 & 0 & 0 \\ \sqrt2 & 0 & 0 \end{array}\right)\;.
\end{equation}

\vspace{0.2cm}

\ni Due to Eqs. (16), $a$ and $a^\dagger $ change the eigenvalues $n_i \!=\! 0,1,2$ (or $N_i \!=\! 1,3,5)$ of the matrix $n$ (or $N$) by $-1$ and +1 (or $-2$ and +2), respectively, {\it within} the range 0 to 2 (or 1 to 5){\footnote{Three generations $i = 1,2,3$ may be also labelled by $n_i = 0,1,2$ or by $N_i = 1 + 2n_i = 1,3,5$. In the model of three fundamental-fermion generations based on the generalized Dirac equation proposed some years ago [5, 4], the label $N_i$ is the number of bispinor indices which appear in three generalized Dirac wave functions describing fundamental fermions of three generations. It is conjectured that among these $N_i$ bispinor indices ("algebraic partons") there are $N_i - 1$ undistinguishable and antisymmetrized (so, obeying "Pauli principle realized intrinsically"), while there is only one distinguished by its coupling to the \SM gauge interactions. The antisymmetrized bispinor indices appear in pairs: the label $n_i = (N_i -1)/2$ is the number of such pairs present in three generations. The number $N_i - 1$ of antisymmetrized bispinor indices, as taking four values each, cannot exceed four. Thus, we conclude that $N_i$ is necessarily equal to 1 or 3 or 5, what explains the existence of exactly three generations of fundamental fermions.}}.

In the formalism of $a$ and $a^\dagger $, it is natural to conjecture tentatively that the active-neutrino mass matrix has the form $ M = (M_{\alpha \beta})$ with $M_{\alpha \beta}$ as given in Eqs. (4) and (7), {\it but} its off-diagonal part can be presented -- more restrictively -- in terms of our annihilation and creation matrices in the following way:
 
\vspace{0.1cm}

%rownanie 19
\begin{eqnarray} 
\left( \begin{array}{rrr} 0\:\: & M_{e\,\mu} & -M_{e\,\mu} \\ M_{e\,\mu}  & 0\:\: & M_{\mu\,\tau} \\ 
-M_{e\,\mu}  & M_{\mu\,\tau} & 0\:\: \end{array}\right) & = & \mu \,\rho^{1/2} \left[g(a + a^\dagger) - g'(a^2 + a^{\dagger\,2})\right] \rho^{1/2} \nonumber \\ & & \nonumber \\ 
& = & \frac{\mu}{29} \left( \begin{array}{ccc} 0 & 2g & \!\!\!- 4 \sqrt3 \,g' \\ 2g & 0 & 8\sqrt3 \,g \\ - 4 \sqrt3 \,g' & 8\sqrt3 \,g & 0 \end{array}\right)\;\,, 
\end{eqnarray}
 
\vspace{0.2cm}

\ni where $g>0$ and $g' >0$ are free parameters (multiplied by the mass scale $\mu$ introduced in Eqs. (7)), while

%rownanie 20
\begin{equation} 
\rho^{1/2} = \left( \begin{array}{ccc} \rho^{1/2}_1 & 0 & 0 \\ 0 & \rho^{1/2}_2 & 0 \\ 0 & 0 & \rho^{1/2}_3 \end{array}\right) = \frac{1}{\sqrt{29}} \left( \begin{array}{ccc} 1 & 0 & 0 \\ 0 & \sqrt4 & 0 \\  0 & 0 & \sqrt{24} \end{array}\right)\,.  
\end{equation}
 
\vspace{0.2cm}

\ni Here, $\rho_i $ are the generation-weighting factors defined in Eqs. (9) in the context of mass formula (6). Consequently, they ought to appear also in the off-diagonal part (19) of mass matrix $M$. We can see from Eqs. (4) and  the equality (19) that

% rownanie 21
\begin{equation}
-\frac{1}{3}(m_1 - m_2)= M_{e\,\mu} =\frac{\mu}{29} 2 g = \frac{\mu}{29} 4\sqrt3 g'
\end{equation}

\ni and 
  
% rownanie 22
\begin{equation}
-\frac{1}{6}(m_1 +2m_2 - 3 m_3)= M_{\mu\,\tau} =\frac{\mu}{29} 8\sqrt3 \,g = 4\sqrt3 \,M_{e\,\mu} = -\frac{4}{\sqrt3}(m_1 - m_2) \;.
\end{equation}

Thus{\footnote{Note that the relations $g = 2\sqrt{3}\, g'$ and $M_{\mu \tau} = 4\sqrt{3}\, M_{e \mu}$ following from the conjecture (19) are valid more generally for the bilarge mixing matrix $U$, where $c_{23} = 1/\sqrt2 = s_{23}$ and $s_{13} = 0$, while the angle $\theta_{12}$ in $c_{12}$ and $s_{12}$ is a free parameter determined from the experimetal data. Then, $0< g = 2\sqrt3 \,g' = (29/\sqrt{8} \,\mu)c_{12}s_{12}(m_2 - m_1)$ and the coefficient $\eta $ in Eq. (24) is $\eta \equiv c^2_{12}(1 + 8\sqrt{3/2} \,t_{12})$ with $t_{12} \equiv  \tan \theta_{12}$.}}, Eq. (21) determines $g$ and $g'$ through $m_2 - m_1$:

% rownanie 23
\begin{equation}
0 < g  = 2\sqrt3 \,g' = \frac{29}{6\mu} \,(m_2 - m_1) \;,
\end{equation}

\ni and Eq. (22) {\it predicts} $m_3$ in terms of $m_1$ and $m_2$:
 
%rownanie 24
\begin{equation}
m_3  =  \eta \,m_2 - (\eta - 1) \,m_1 =  \eta(m_2 - m_1) + m_1
\end{equation}
 
\ni with
 
%rownanie 25
\begin{equation}
\eta \equiv \frac{2}{3} \,(4\sqrt3 +1) = 5.28547\,.
\end{equation}

\ni Notice that Eq. (24) allows in particular for the limiting option of exact degeneracy $ m_1 =  m_2 = m_3$ that is excluded experimentally. Generically, from the restrictive relation (24) we get the equation
 
%rownanie 28
\begin{equation}
[\eta \,m_2 - (\eta - 1) m_1]^2 - m^2_2 = m^2_3 - m^2_2  =  \lambda(m^2_2 - m^2_1) 
\end{equation}
 
\vspace{-0.2cm}

\ni or
 
\vspace{-0.1cm}

%rownanie 27
\begin{equation}
[\eta - (\eta - 1) r]^2 - 1 -   \lambda(1 - r^2) = 0 \,,
\end{equation}
 
\vspace{-0.2cm}
 
\ni where
 
\vspace{-0.1cm}
 
%rownanie 28
\begin{equation}
r \equiv \frac{m_1}{m_2}
\end{equation}
 
\vspace{-0.2cm}

\ni and
 
\vspace{-0.1cm}
 
%rownanie 29
\begin{equation}
\lambda \equiv \frac{m^2_3 - m^2_2}{m^2_2 - m^2_1} \sim \frac{2.4\times 10^{-3}}{8.0\times 10^{-5}} = 30 \,,
\end{equation}

\vfill\eject

\ni the latter value being valid in the case of popular experimental best fit [2] (here 
$m_1 < m_2 < m_3$).

With $\lambda > \eta - 1 = 4.28547$, Eq. (27) for $r$ gets two solutions

%rownanie 30
\begin{equation}
r^{(1,2)} = \frac{\eta(\eta -1)}{(\eta -1)^2 +\lambda}  \pm \frac{|\eta -1 - \lambda |}{(\eta -1)^2 +\lambda} = \left\{ \begin{array}{l} 1 \\ \frac{\eta^2 -1 -\lambda}{(\eta -1)^2 +\lambda} \end{array} \right.\,.
\end{equation}

\vspace{0.2cm}
  
\ni The first solution $r^{(1)} = 1$ corresponds to the limiting option of exact neutrino mass degeneracy $ m_1 = m_2 = m_3$, what is not the case realized experimentally. The second solution is nonnegative, $r^{(2)} \geq 0$ (giving $m_1 \geq 0$), {\it only} if $\lambda \leq \eta^2 - 1 = 26.9362$ (while the popular experimental best fit is $\lambda = 30$). With the use of experimental estimate $m^2_2 - m^2_1 \sim 8.0\times 10^{-5}\;{\rm eV}^2$, this solution gives $m^2_3 - m^2_2 \stackrel{<}{\sim} 2.2\times 10^{-3}\;{\rm eV}^2$.

For the maximal allowed value  $\lambda = \eta^2 - 1 = 26.9362$ we get $r^{(2)} = 0$ and so, we {\it predict} $m_1 = 0$ and $m^2_3 - m^2_2 \sim 2.2\times 10^{-3}\;{\rm eV}^2$ (while the popular experimental estimate is $m^2_3 - m^2_2 \sim 2.4\times 10^{-3}\;{\rm eV}^2$). Such a smaller value is still consistent with the present data within experimental limits. With $m_1 = 0$, we now obtain 

%rownanie 31
\begin{equation}
m_2 \sim 8.9\times 10^{-3} \;{\rm eV}\;,\;  m_3 \sim 4.7\times 10^{-2} \;{\rm eV}
\end{equation}
 
\ni and infer from Eq. (23) that 
 
%rownanie 32
\begin{equation}
g = 2\sqrt{3} \,g' \sim 0.53\,.
\end{equation}

\ni In the last estimation, the value $\mu \sim 8.1\times 10^{-2} \;{\rm eV}$ (Eq.(33)) is applied. From Eq. (31) we can determine with $m_1 =0$ the following parameter values in Eq. (6) in place of the previous values (13):

%rownanie 33
\begin{equation}
\mu \sim 8.1\times 10^{-2} \;{\rm eV}\;,\;  \frac{\varepsilon}{\xi} = 1 \;,\;  \frac{1}{\xi} \sim 1.0\times 10^{-2} \;.
\end{equation}

\vspace{0.2cm}

\ni Here, the parameter $1/\xi $ in Eq. (6) is still small {\it versus} 1 and $\varepsilon/\xi $, though it is larger than previously.

Concluding, we have constructed in this note the off-diagonal part of the active-neutrino mass matrix with the use of two $3\times 3$ matrices playing the role of annihilation and creation matrices acting in the neutrino-generation space of $ \nu_e \,,\, \nu_\mu \,,\, \nu_\tau $. The construction leads to the new relation $M_{\mu \tau} = 4\sqrt{3}\, M_{e \mu}$ (Eq. (22)) which {\it predicts} in the case of tribimaximal neutrino mixing that $m_3 - m_1 = \eta(m_2 - m_1)$ with $\eta = 5.28547$. Then, the maximal possible value of $\lambda \equiv (m^2_3 - m^2_2)/(m^2_2 - m^2_1)$ is equal to $\eta^2 - 1 = 26.9362$ and gives $m_1 = 0$. With the experimental estimate $m^2_2 - m^2_1 \sim 8.0\times 10^{-5}\;{\rm eV}^2 $, this maximal value, if realized, {\it predicts}  $m^2_3 - m^2_2 \sim 2.2\times 10^{-3} \;{\rm eV}^2$, near to the popular experimental estimation $m^2_3 - m^2_2 \sim 2.4\times 10^{-3} \;{\rm eV}^2$.

Formally, one may say that our neutrino mass formula (6), which primarily is a special transformation of three masses $m_1, m_2, m_3$ into three parameters $\mu, \varepsilon, \xi $, becomes a special function predicting $m_1=0$, $m_2$ and $m_3 = \eta \,m_2$ (with $\eta = 5.28547$) in terms of $\mu$ and $\varepsilon = \xi $, when the conjecture (19) about the 
off-diagonal part of neutrino mass matrix is made and the value $\lambda \equiv (m^2_3 - m^2_2)/({m^2_2 - m^2_1}) \leq \eta^2 -1$ assumed to be maximal: $\eta^2 -1$. More generally, without the latter requirement, still $m_3 = \eta \,m_2 - (\eta -1)m_1$ is predicted in terms of $m_1$ and $m_2$, what imposes a relation between $\varepsilon $ and $\xi $.

\vfill\eject

~~~~
\vspace{0.5cm}

{\centerline{\bf References}}

\vspace{0.5cm}

{\everypar={\hangindent=0.6truecm}
\parindent=0pt\frenchspacing

{\everypar={\hangindent=0.6truecm}
\parindent=0pt\frenchspacing

[1]~L. Wolfenstein,  {\it Phys. Rev.} {\bf D 18}, 958 (1978); P.F. Harrison, D.H. Perkins and W.G.~Scott, {\it Phys. Lett.} {\bf B 458}, 79 (1999); {\it Phys. Lett.} {\bf B 530}, 167 (2002); Z.Z.~Xing. {\it Phys. Lett.} {\bf B 533}, 85 (2002); P.F. Harrison, and W.G.~Scott, {\it Phys. Lett.} {\bf B 535}, 163 (2003); T.D.~Lee, {\tt hep--ph/0605017}; {\it cf.} also W. Kr\'{o}likowski,  
{\tt hep--ph/0509184}.

\vspace{0.2cm}

[2]~{\it Cf. e.g.} G.L. Fogli, E. Lisi, A. Marrone and A. Palazzo, {\tt hep--ph/0506083}.

\vspace{0.2cm}

[3]~ W. Kr\'{o}likowski, {\tt hep--ph/0602018}.

\vspace{0.2cm}

[4]~W. Kr\'{o}likowski, {\tt hep--ph/0604148}.

\vspace{0.2cm}

[5]~W. Kr\'{o}likowski, {\it Acta Phys. Pol.} {\bf B 32}, 2961 (2001) [{\tt hep-ph/0108157}], Appendix; {\bf B~33}, 2559 (2002) [{\tt hep-ph/0203107}]; and earlier references therein.

\vfill\eject

\end{document}